# Roadblocks to Attracting Students to Software Testing Careers: Comparisons of Replicated Studies

Rodrigo E. C. Souza[1], Ronnie E. de Souza Santos[1,2], Luiz Fernando Capretz[3], Marlon A. S. de Sousa[1] and Cleyton V. C. de Magalhães[1]

[1] Agile Testing Program, CESAR School, Recife – Pernambuco - Brazil
[2] Faculty of Computer Science, Dalhousie University, Halifax – Nova Scotia – Canada
[3] Dep. of Electrical & Computer Engineering, Western University, London – Ontario – Canada
`recsouza@gmail.com, souzasantos.ronnie@gmail.com,`
`lcapretz@uwo.ca, mass@cesar.school, cleyton.vanut@gmail.com`

**Abstract. Context**. Recently, a family of studies highlighted the unpopularity of software testing careers among undergraduate students in software engineering and computer science courses. The original study and its replications explored the perception of students in universities in four countries (Canada, China, India, and Malaysia), and indicated that most students do not consider a career in software testing as an option after graduation. This scenario represents a problem for the software industry since the lack of skilled testing professionals might decrease the quality of software products and increase the number of unsuccessful projects. **Goal**. The present study aims to replicate, in Brazil, the studies conducted in the other four countries to establish comparisons and support the development of strategies to improve the visibility and importance of software testing among undergraduate students across the globe. **Method**. We followed the same protocol in the original study to collect data using a questionnaire and analyzed the answers using descriptive statistics and qualitative data analysis. **Results**. Our findings indicate similarities among the results obtained in Brazil in comparison to those obtained from other countries. We observed that students are not motivated to follow a testing career in the software industry based on a belief that testing activities lack challenges and opportunities for continuous learning. **Conclusions**. In summary, students seem to be interested in learning more about software testing. However, the lack of discussions about the theme in software development courses, as well as the limited offer of courses focused on software quality at the university level reduce the visibility of this area, which causes a decrease in the interest in this career.

## 1   Introduction

Testing is an indispensable part of software development. It is the activity responsible for verifying that a system meets the planned requirements and validating that it satisfies its intended purpose [1]. The history of software development indicates that test-ing activities existed before the establishment of processes, practices, and models for software development [2] [3], which reinforces the importance of this area. The relevance of software testing has been studied and practiced since the beginning of computing as researchers and practitioners are consistently following the way this activity evolved from a simple task focused on checking the results obtained from the source code execution to a leading and interactive process essential for the development of software products [2]. As a result, especially given the agile nature of software development, testing activities are widely spread throughout the development process and the system is continuously tested [1] [2] [3].

Precisely because testing has proven to be a vital activity in software development, researchers and practitioners are frequently searching for different approaches and techniques to improve the testing process and the resulting software quality. Among other strategies, such improvement could be reached by increasing the number of joint industry-academia collaborations [4]. However, recent studies have indicated that researchers and practitioners are not collaborating enough to solve industrial problems [5] [6] [7] [8]. The reality is that the distance between industry and academia likely existed before scientific and research contexts, as careers focused on software testing appear to be underrated by undergraduate students in software and computer engineering programs [9]. The unpopularity of software testing among students is pointed to as the main finding of a family of replications recently published in two studies that investigated the perception of students in universities across four countries, namely, Canada, China, India, and Malaysia [10] [11].

These findings have direct implications for academia and industry since they demonstrate the need for improving the perception of testing careers among students to prepare highly skilled professionals to work in this area in software companies. However, since the results are dependent on cultural and social factors related to the country where the data was collected, more replications would be the appropriate next step for extending discussions and increasing generalizability. The lack of skilled testing professionals is a major issue for the software industry because of the centrality of core quality elements to a successful project. In addition, usually testing processes take up about 40-50% of a project's time [5], which produces a direct impact on costs and deliveries. Therefore, understanding the interests of students from several different regions in software testing, as well as their reasons for taking up or not testing careers is an aspect of software engineering that requires immediate attention. In this sense, the present study is a replication of the above-cited studies [10] [11] focused on answering the following general research question:

*RQ: How do Brazilian undergraduate students from software and computer engineering perceive software testing careers?*

In this paper, we present the initial results obtained by utilizing the same survey used in [10] and [11], and this introduction, is organized as follows. In Section 2, we present a brief background about [10] [11] hereafter referred to as original study and first replication, respectively. In Section 3, we describe how this replication was conducted and how the research method was applied. In Section 4, we present the main findings and discussions. Finally, in Section 5, we present our conclusions and directions for future investigations.

## 2    Background

This section discusses findings from the literature on software testing in software and computer engineering programs and presents the original study and the first replication, which were replicated in this research.

### 2.1    Software Testing in Academic Curricula

The way students are trained in the academy including the content they study, but also their experiences in class—impacts their development as professionals. In other words, the professional education that individuals receive before joining any organization will have a direct effect on their perception of the work, and consequently, on what profession they will opt to take up [12].

When young software professionals start to work in the industry, they will depend almost exclusively on what they have learned at the university over the previous four or five years. However, recent studies have emphasized the existence of a gap between the software engineering industry and software engineering education. Such a gap arises from a series of elements, such as a lack of activities to develop soft skills, and considerable differences between projects developed at the university and real-life industry projects (e.g., size, requirement details, management, etc.), which leads to a third element, the distance of the school from the actual industry reality [13] [14].

In this scenario, the role of software testing in software and computer engineering programs is curious. Since the early 2000's researchers have noted that this area received little treatment in most curricula, even though it can represent almost 50% of the cost of software [15]. Over the years, as quality became essential, companies started to indicate that students should develop good problem solving, debugging, and analysis skills, since many graduates begin their careers in industry with exceptional programming skills, but lack competence in testing, debugging, and analysis skills [16] [17].

To address this problem, we need to introduce and improve teaching in undergraduate software testing programs. This will have the goal of enabling students to recognize the importance of testing and quality in software development, while also solving practical problems by applying contemporary technologies and methods to verification and validation processes [16] [18]. However, a recent study reported an additional new challenge regarding this difficulty, the unpopularity of testing careers among software practitioners [19].

## 2.2 Replications of Empirical Studies in Software Engineering

The replication of empirical studies represents an important component in the construction of knowledge in software engineering. Through replications, studies can be repeated, results can be checked, and the validity of outcomes can be expanded to different contexts [21].

In software engineering, replications are mostly used to generalize the results of an original study to a different population [24]. According to [25] other uses for replications in software engineering include:

a)  confrontation of results from a new study in contrast to previous ones.
b)  improve the research design of a previous study.
c)  increase the external validity of results from previous investigations.
d)  improvement research skills.
e)  understand costs and efforts for future studies.

According to this definition, the main goal of the present research is related to (a) and (c), since we replicated a study conducted in four different countries to check how the findings apply to a fifth one.

## 2.3 Original Study and First Replication

The present study is a replication of two previous research papers conducted in Canada, China, and India (original study) [10], and posteriorly in Malaysia (first replication) [11]. Both research papers aimed to investigate the perception of undergraduate students in software and computer engineering programs of a career in software testing to discuss the (un)popularity of this profession [19]. Based on the classification of replications in software engineering, we consider the present study as an external replication. This means that the replication was performed by a different group of researchers [24].

In all three studies, the research method conducted to address this problem was a survey, which was designed to collect the opinion of several undergraduate students by applying a questionnaire to answer three main questions:

a)  What is the likelihood of them taking up a career in software testing?
b)  What are the advantages of taking up a career in software testing?
c)  What are the drawbacks of taking up a career in software testing?

For the first question, the participants selected one of the provided options, namely, *Certainly Yes*, *Yes*, *Maybe*, *No*, and *Certainly Not*. The following two questions were open-ended. For data analysis, the authors pointed out that a qualitative approach was applied to explore the phenomena within their real-life context.

The original study obtained answers from 254 computer and software engineering students from three different countries, 85 participants from Canada, 99 participants from China, and 70 participants from India. Following this study, the first replication surveyed 82 students from software engineering-related programs, such as information technology and computer science at two Malaysian universities. The general results demonstrated that software testing is very unpopular, especially among stu-dents from Canada, China, and India, while a career in this area would be considered by an average number of Malaysian students. Table 1 summarizes these results, which will be used to discuss the results obtained from the current replication.

**Table 1.** Choosing a Career in Testing

| Responses | Canada | China | India | Malaysia |
|---|---|---|---|---|
| Certainly No | 31% | 24% | 14% | 1% |
| No | 27% | 0% | 31% | 7% |
| Maybe | 33% | 74% | 47% | 52% |
| Yes | 7% | 2% | 7% | 34% |
| Certainly Yes | 2% | 0% | 0% | 6% |

Regarding the advantages of working with software testing, the original study and the first replication demonstrated that viewing this career as a learning opportunity and as comprising easy tasks were the main benefits observed across Canada, China, India, and Malaysia, although the percentage varies considerably among the four countries. The number of positions available caught the attention of Canadian, Chinese, and Malaysian students, while Canadian, Indian, and Malaysian students consider software testing an important job, which represents an advantage. Other benefits highlighted by the participants include monetary benefits and fun during work, e.g., exploring and finding bugs.

On the other hand, the drawbacks associated with the work in software testing are the monotony, which is present across the four countries, and the complexity, which is less perceived by Canadian students. The lack of development activities is also a disadvantage pointed out by all the groups of students. Other drawbacks include a lack of interest, especially in finding others' mistakes (code mistakes), and the lack of recognition in the industry. Minor disadvantages would be related to low salary in comparison to other professionals and stressful activities.

In summary, the results obtained from the original study and the first replication demonstrate that the perception of students about a career in software testing varies significantly depending on the country, which will require specific and targeted actions to emphasize the importance of testing activities to undergraduates and to highlight the perks of working with software testing in the industry. In addition, the findings revealed the existence of myths among students, such as the belief that the testing process always lacks programming.

We expect that our replication represents a step forward in improving the knowledge acquired so far. Thus, based on the data collected from five countries and over 400 students we will be able to start designing and proposing strategies to improve the popularity of software testing in the academy, and consequently increase the number of highly skilled professionals to work in this area in the software industry.

Lastly, even though we are replicating two previous studies that are interrelated, e.g., the original study [10] and the first replication [11], it is important to mention that additional studies focused on this theme may be available in the literature, and these can be analyzed in the future to improve the results of the current research. As an example, Deak et al. [20] investigated the factors that influence Norwegian students when deciding to choose a career in the area of software testing and based on the results identified strategies that can be used to motivate students and improve course contents.

## 3        Method

In this study, we consider replication as a conscious and systematic repeat of an original study [22]. Therefore, we followed the same protocol used by the original study and the first replication to collect and analyze data, as described below.

### 3.1        Data Collection

Following the previous studies, in the present replication, we applied the same questionnaire to collect data from undergraduate students from software and computer engineering-related programs. However, the instrument was slightly modified to achieve the goal of our research. First, the questionnaire was translated into the native language of the targeted participants (i.e., Portuguese). Second, we introduced the questionnaire with a quick definition of software testing, so all respondents would have basic knowledge of the topic under study. Third, we added to the questionnaire a quick question asking about the undergrad level of the respondents, e.g., what year of the undergraduate program the students are enrolled in. Lastly, we added an extra question asking students to justify why they would consider or not consider a career in software testing. We believe that more qualitative data associated with this closed question presented in the previous studies could help in the process of proposing solutions to the main problem observed in this context, e.g., motivating young professionals to work with software quality in the industry.

The translated questionnaire was validated through a pilot round, conducted with five members of our research group, which are not involved in this study. They were asked to read and compare the original questions with the translations, answer the questionnaire, and provide feedback about it. No update was performed after the pilot round since the questions are straightforward. Table 2 presents the final version of the questionnaire applied in this study. Once the instrument was validated the research team started to announce the research and the questionnaire to student groups, professors, researchers, and professionals, asking for help to collect the data from the targeted population.

Regarding the population, our study focused on all students enrolled in software/computer engineering programs and popular related programs in this area in Brazil such as information systems, computer science, and technology, among others. Invitations were sent to all regions of the country and no restriction was defined regarding the student level, which means that we were expecting data from individuals that were starting at the university right up to those about to graduate. Data collection ran for about two months and all questionnaires received were anonymous.

**Table 2.** Questionnaire

According to the SWEBOK, software testing is defined as the dynamic process of verifying and validating a software under development to attest it works as expected and possesses all the planned features and behaviors. Based on this definition and your previous knowledge/experience with software testing before your graduation, please answer the following questions.

| Topic | Questions |
|---|---|
| **Choosing a career in software testing** | 1. After you graduate, would you consider a career in software testing?<br>( ) Certainly No<br>( ) No<br>( ) Maybe<br>( ) Yes<br>( ) Certainly Yes<br>2. Please, briefly justify our answer for the previous question. |
| **Advantages and Drawbacks** | 3. What are three advantages (from the most to less important) of taking up a career in software testing?<br>4. What are three drawbacks (from the most to the less important) of taking up a career in software testing? |

### 3.2 Data Analysis

The nature of the questions posed to participants in this study required both quantitative and qualitative approaches to data analysis. Descriptive statistics were applied to analyze the answers to closed-ended questions designed to assess the likelihood of students deciding to work in software testing. Following this, qualitative analysis was applied to consolidate, reduce, and interpret all data obtained from open-ended questions, which were focused on revealing the advantages and drawbacks of software testing from the participants' perspective, along with a descriptive answer to the close-ended question.

The guidelines for conducting qualitative research suggest that the process should be based on coding the answers provided by participants and making sense of them [23]. In qualitative analysis, open coding is the process of reducing the narratives collected in interviews or questionnaires into discrete parts which can be closely examined and compared, looking for similarities and differences, and organizing concepts into representative categories [23]. This is the main process followed in this research to synthesize the answers collected from students. Figure 1 illustrates the open coding process and the construction of categories developed in this study.

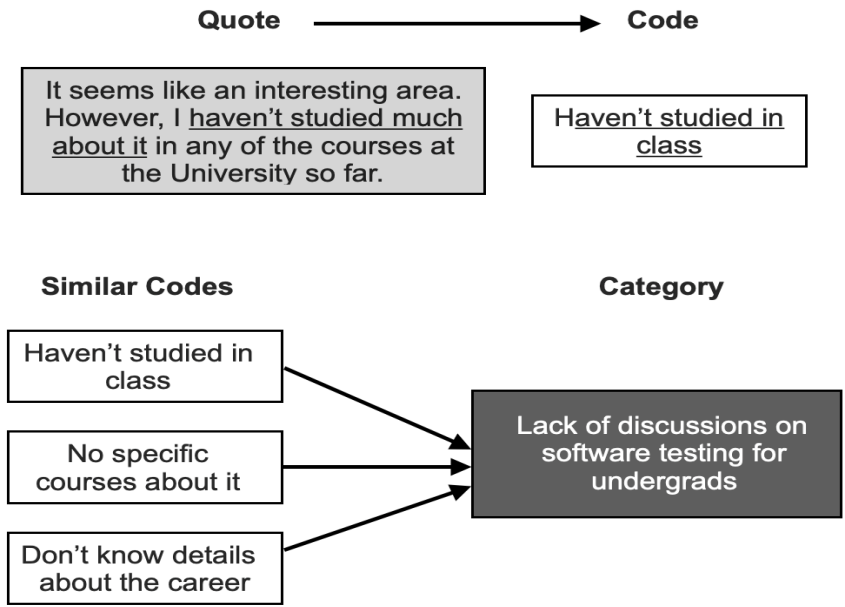

**Fig. 1.** Qualitative analysis: open coding.

## 4 Findings

We obtained 92 valid questionnaires with answers from Brazilian students, distributed as follows: 29% of students were in their first year, 32% in their second year, 16% in their third year, 13% in the fourth, and 10% in the fifth year. Unlike in other countries, it is common in Brazil for students to take up to 5 years of instruction in colleges and universities. However, in computer/software engineering programs the dropout rate tends to increase after the second year. This is one of the factors that explains the larger number of participants in the first and second years of this study. Following this general characterization of our sample, we answer each of the questions presented on the questionnaire.

### 4.1 After you graduate, would you consider a career in software testing?

Initial results indicate asymmetry in the perception of Brazilian students regarding the popularity of working with software testing since the likelihood of respondents choosing to work in this area is demonstrated to be well-balanced. About 27% of students expressed an interest in taking up this career. The same percentage of respondents expressed no desire to work in this area whatsoever. On the other hand, almost half of the sample (46%) indicated that they could (or could not) choose this career by answering *maybe* to this question. Figure 2 summarizes this information.

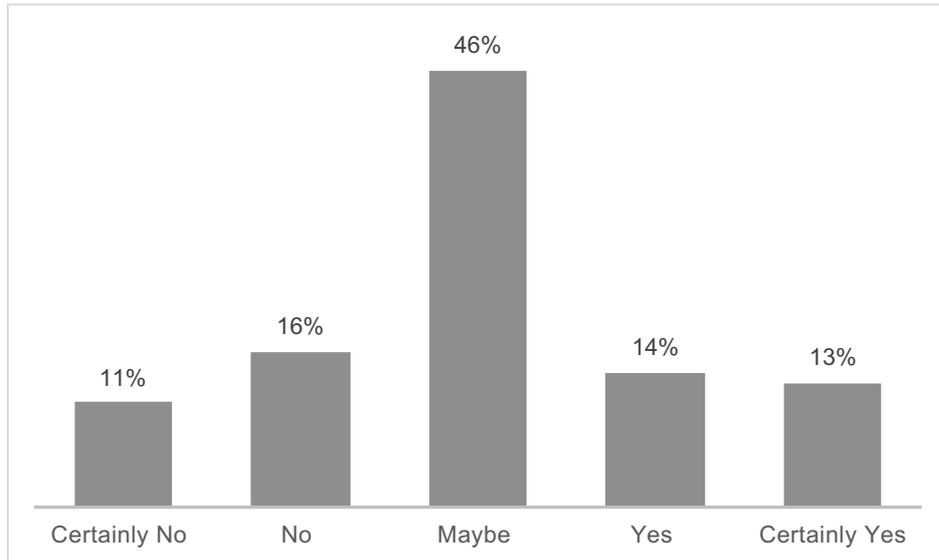

**Fig. 2.** Likelihood of Brazilian Students Choosing a Career in Software Testing

By comparing the results from the present study with the results from the original study and the first replication, we observed that considering the countries surveyed so far, work in software testing is more popular in Malaysia, followed by Brazil, then, Canada, India, and China at lowest. Table 3 summarizes these results.

**Table 3.** Choosing a Career in Testing - Second Replication

| Responses | Brazil | Canada | China | India | Malaysia |
|---|---|---|---|---|---|
| Certainly No | 11% | 31% | 24% | 14% | 1% |
| No | 16% | 27% | 0% | 31% | 7% |
| Maybe | 46% | 33% | 74% | 47% | 52% |
| Yes | 14% | 7% | 2% | 7% | 34% |
| Certainly Yes | 13% | 2% | 0% | 0% | 6% |

However, both Malaysia and Brazil present a similar outcome regarding the percentage of students who would be inclined to work in software testing, which are those individuals who answered *maybe*. Many factors could explain this reality. However, such explanations are outside of the scope of this study at this point. We can only hypothesize, based on the literature, that cultural aspects of each country and the dynamics of university programs might be core factors influencing this reality. Further, unlike the previous studies, our study also requested participants to justify their answers to this question. Thus, by applying open coding we obtained the main broad reasons students cited for being willing or unwilling to consider a career focused on testing.

There are two main reasons associated with the fact that 27% of students would not work with software testing after graduation (Certainly No, and No responses). First, some individuals in this group have already developed interests in other areas of software development and expect to work in those areas in the future. Second, some students have an outdated perception about the job and the impact of software testing on software development, e.g., individuals relate testing with a lack of opportunity for coding and monotony at work, while in other cases they cannot even perceive the connection between testing and the rest of the software development activities.

Those who answered *Certainly Yes* and *Yes* to the possibility of working with software testing in the future seem to be oriented by the previous contact with the area. First, some students who attended courses or lectures focused on software testing, developed a positive attitude regarding this career, which turned into a willingness to experiment more in this area. Second, some students first interacted with software testing through internships and now they want to continue working in this area. Finally, some students do not have software testing as their main interest, but they are open to working with it depending on several factors, such as payment, benefits, and learning opportunities, among others.

In our analysis, the reasons for the *Maybe* response proved to be more dynamic and fluid than the *Yes/No* answers. Therefore, this analysis will be part of our future work. Table 4 presents quotations extracted from the questionnaires that elucidate the students' reasons.

**Table 4.** Students' reasons for choosing a career in testing

| After you graduate, would you consider a career in software testing? | |
|---|---|
| **Answer** | **Justification (quotations)** |
| **Certainly No / No** | - "I am not an enthusiastic of software testing. I rather be coding new features". <br> - "Certainly, never caught my attention". <br> - "I want to create games in the future, not this thing." <br> - "Testing is not really of my taste. I like challenges and working with new people, new problems…" |
| **Certainly Yes / Yes** | - "Since I am working with this lately, I don't really see me working with something else [after graduation]. This is what I like". <br> - "I liked it, since I learnt about it on a lecture." <br> - "Testing is an area that caught my attention, it will give opportunity to learn a lot". |

## 4.2 What are the advantages and drawbacks of taking up a career in software testing?

Similar to what was pointed out in the previous paragraph, in this paper, the analysis of advantages and drawbacks pointed out by students is still underway. Based on the data, it is possible to indicate the most often cited advantages and drawbacks. However, further analysis is necessary to provide a representative description for each of these elements, considering not only their meaning but also their relationship with several factors, e.g., the student's level (years), their attitude towards software testing (more positive or negative), their previous experience with testing, e.g., through internships, among others.

We identified 28 advantages of working with software testing based on participant responses. In the sample, the most prominent benefits are:

a) The number job of positions and opportunities currently available for professionals was cited by 39% of individuals.

b) Payment and financial compensation for professionals in this career were cited by 38% of individuals.

c) The sense of satisfaction in supporting the release of the software with high levels of quality was cited by 28% of individuals.

d) Constant learning and training opportunities were cited by 17% of the subjects.

e) Challenges at work were cited by 16% of individuals.

The list includes other advantages such as the possibility of supporting other areas of software development, working with programming (test automation), do not work with programming, and teamwork, among others. On the other hand, we obtained 32 different drawbacks cited by students regarding the work with software testing. The most representative disadvantages of this profession would be:

a) The fact that the work is monotonous and not interesting enough was mentioned by 30% of the participants.

b) The repetitiveness of tasks was cited by 22% of participants.

c) The work is stressful, according to 16% of participants.

d) The salary is low in comparison to other professionals, according to 15% of participants.

In addition, 15% of participants claim to have e no knowledge about the area, which would explain several drawbacks. The list of disadvantages also includes a lack of opportunity for coding activities, complexity, low number of positions available, and low relevance for software development as software testing career.

## 5 Conclusions

We presented the results obtained from the replication of studies that were conducted in Canada, China, India, Malaysia, and Brazil. In summary, our analysis demonstrates similarities among the results obtained in Brazil, India, and Malaysia regarding the perception of undergrad students towards working with software testing. In these countries, students tend to be more receptive and enthusiastic about testing careers. Further analysis can reveal aspects related to these places, e.g., cultural, educational, economic, or technical, that can be used to discuss strategies to improve the visibility of software testing for students in other countries.

Future works include additional analysis of the supplementary qualitative data that we collected from Brazilian students, which can be used to further explore and describe the scenario in Brazil while raising detailed comparisons among the countries researched so far. Long-term future work includes replicating this study with professionals working in the industry to draw a line between expectations (students' perception) and reality (practitioners' routine) regarding the advantages and drawbacks of a software testing career.

This study has implications for both the academy and the software engineering profession. For academia, the comparisons established among the replications might be used to create strategies for improving software/computer engineering programs by including more testing courses to provide students with the knowledge and the skills necessary to work in software testing careers in the industry. For industry, these results create awareness of the need of developing strategies to motivate and engage software QA professionals, in particular trainees and individuals at the beginning of their career. Both strategies are crucial for the development of high-quality software.